\documentclass[hidelinks]{article}
\title{Living Cognitive Society:\\
a `digital' World of Views}
\usepackage[backend=bibtex,natbib=true,style=authoryear-ibid,maxcitenames=2,maxnames=2,url=false,doi=false,isbn=false]{biblatex}
\usepackage[english]{babel}
\usepackage{csquotes}
\usepackage{hyperref}
\usepackage{float}
\hypersetup{
	colorlinks = true,
    linkcolor = blue,
    urlcolor  = blue,
    citecolor = blue,
    anchorcolor = blue]
}
\usepackage[font=scriptsize,labelfont=bf]{caption}

\usepackage{mdframed}

\usepackage{paralist}

\usepackage{graphicx}
\usepackage[disable,textsize=footnotesize]{todonotes}

\author{Viktoras Veitas \texttt{vveitas@gmail.com} \\
David Weinbaum (Weaver) \texttt{space9weaver@gmail.com}
}

\bibliography{../biblio/article_specific,../biblio/vveitas_library}

\begin{document}

\maketitle

\begin{abstract}
The current social reality is characterized by all-encompassing change, which disrupts existing social structures at all levels. Yet the prevailing view of society is based on the ontological primacy of stable hierarchical structures, which is no longer adequate.

We propose a conceptual framework for thinking about a dynamically changing social system: the Living Cognitive Society. Importantly, we show how it follows from a much broader philosophical framework, guided by the theory of individuation, which emphasizes the importance of relationships and interactive processes in the evolution of a system.

The framework addresses society as a living cognitive system -- an ecology of interacting social subsystems -- each of which is also a living cognitive system. We argue that this approach can help us to conceive sustainable social systems that will thrive in the circumstances of accelerating change. The Living Cognitive Society is explained in terms of its fluid structure, dynamics and the mechanisms at work. We then discuss the disruptive effects of Information and Communication Technologies on the mechanisms at work.

We conclude by delineating a major topic for future research -- distributed social governance -- which focuses on processes of coordination rather than on stable structures within global society.
\end{abstract}

\textbf{Keywords}: cognitive system, living society, information and communication technologies, future social governance, individuation, cognitive development.

\section{Introduction}\label{introduction}

Today's society and life in general are characterized by 
all-encompassing fast change and movement. New technologies, new jobs,
new opportunities, new challenges -- i.e.~new unknowns -- seem to fall on
us before we can figure out how to make sense of the current
ones. Our psychological reactions vary between: (1) attempts to
`stabilize' the environment (social, political, technological,
biological) by imposing more controls and checkpoints; (2) calls to
embrace change and ride its wave towards a `new world order'; (3)
\textit{ad-hoc} proposals for dealing with challenges of our times
(e.g.~information overload); or (4) a sense of helpless
dis-attachment.

No matter what the specific reaction is to the socio-technological
change we are experiencing, it is based on the way in which we make sense of
ourselves, others and the world. Usually we base our sense-making on
perceivable stable objects and their relationships in the world. A
specific configuration of such objects and relationships within a system
describes its state. The change of the system is then perceived as a
chain of transitions between states. This is a well-established mode of
thinking which has helped us tremendously in achieving most of what human
civilization has created since its beginning. But is it still valid in the
era of ever-accelerating change?

This paper proposes an original conceptual framework for thinking about
a changing social system and applies it to the contemporary situation of
the global information society. The gist of the framework is an
approach to a social system as a living cognitive system -- an ecology of
interacting social subsystems.
We do this by developing the concept of
\textit{the Living Cognitive Society} -- a distributed social system
characterized by the interaction of a multiplicity of heterogenous agents
and subsystems. First, we analyse the current situation of global
society and its underlying reasons, and ask the question `what kind of global
system could be sustained and thrive in these circumstances?' (Section
\ref{the-current-situation-of-the-global-society}). Then we provide a
detailed tour of the theoretical concepts which form the basis of the
framework (Section \ref{conceptual-background}). The description of the
main concepts is followed by the rationale of their integration (Section
\ref{connecting-the-dots}), which explains the application of the
theoretical basis of our framework to the situation of global
society. The locus of the paper is the detailed characterization of the
Living Cognitive Society in terms of its structure, dynamics and the
mechanisms at work (Section \ref{the-living-cognitive-society}),
building on notions and concepts introduced in the previous sections.
Finally, we apply the concept and mechanisms of the framework to discussion of 
the impact of information and communication technologies
(ICT), particularly the Internet, on global society (Section
\ref{the-disruptive-impact-of-information-and-communication-technologies}).
The issue of the governance of the Living Cognitive Society is intricately
related to the mechanisms at work within the system, and also represents
a distinct challenge and the field of research. We therefore dedicate
the last section to introducing the paradigm of
\textit{distributed governance} (Section
\ref{the-governance-of-fluid-society}) as an avenue for future research.

We aspire to several goals simultaneously with this work. Most
importantly, we aim to construe how the concept of the Living Cognitive
Society, and our approach to the global information society, follow
from a much broader philosophical and theoretical framework, guided by
the theory of individuation. Therefore, while the theoretical framework
alone has been developed elsewhere
\citep{veitas_world_2015,weinbaum_synthetic_2014,weinbaum_framework_2012},
this paper provides an integrated summary of the main concepts with
references to appropriate sources.

Hence, the paper combines: the conceptual framework (Sections
\ref{conceptual-background} and \ref{connecting-the-dots}); the
application of the framework to social reality (Section
\ref{the-living-cognitive-society}); the role of ICT and Internet in the
disruptive change of the global social system (Section
\ref{the-disruptive-impact-of-information-and-communication-technologies});
`connective tissue' -- the interpretation of the current situation of society 
(Section \ref{the-current-situation-of-the-global-society})
and consolidation of concepts with application (Section
\ref{connecting-the-dots}); and finally, avenues for future research (Section
\ref{the-governance-of-fluid-society}). These themes are not 
presented in a linear fashion, but rather intertwined in order to better convey the relation
between the philosophical framework and its application to the global social
system. A number of cross-references are provided in the text for
navigating its thematic structure.

\section{The current situation of global
society}\label{the-current-situation-of-the-global-society}

The current situation of global society can be characterized by the
overwhelming feeling that the world is changing too fast for a single
human and society to comprehend. This feeling furthermore extends to the
inability of coping with change, at least without a paradigmatic
shift in how humans individually and humanity collectively relate to the
world and themselves. There are two aspects to the perception of
disruptive change of our social reality, both playing an important role.
The first is the actual acceleration of the life pace, which can be
connected to the relative, yet increasing, separation of humans from
nature. It is probably rooted in the dawn of human civilization, but
has `become a fully fleshed out experiential concept only with
Industrial Revolution' \citep{hartmut_rosa_is_2009}, and arguably is
reaching its climax with the rise of the `networked world'
\citep{helbing_globally_2013,wef_perspectives_2013}. This separation has
allowed humanity to dissociate its activities from the rhythms of
natural phenomena (day and night, harvesting seasons, etc.) forcing the
socio-technological acceleration on itself. Another aspect is the
psychological reaction to uncertainty, mostly related to 
`information overload' and the `future shock', inherent in our times
\citep{heylighen_change_1999}. Both aspects contribute to the increasing
\textit{social complexity} of our world.

\subsection{Factors of social
complexity}\label{factors-of-social-complexity}

Three major factors of social complexity can be identified: accelerating
change, hyper-connectivity and reflexivity:

\begin{description}
\item[Reflexivity]
is probably the most important characteristic of a social system: it refers to the consideration that the social system is created by the collective behaviour of its participants and, at the same time, exerts an influence on the behaviour on its participants.
Every participant (e.g. person, institution, nation state) of society both affects and is affected by other participants, causing circular internal relationships among them, as well as mutual dependency between participants and the whole society.
Most importantly, reflexivity refers to a feedback relationship between observer / participant of a social system (i.e. intelligent agent) and the observed (i.e. the 'environment' -- the system as a whole).
\item[Hyper-connectivity]
is a major symptom of progress, resulting in a world where every agent, event and process is connected to many other agents, events and processes, therefore making all elements highly interdependent.
The `networked world' is therefore an example of a fragile system, where local events may spread to affect the whole global system (e.g. in the case of stock market crashes).
\item[Accelerating change]
is due to the explosive multiplication of information in the hyper-connected and reflexive system, which is our global information society.
It is a source of uncertainty and confusion in almost all domains of social and human life, because participants of the system have limited capacity to process this information, let alone to match the speed of information multiplication.
\end{description}

The central question which this paper aims to answer is therefore: what
kind of social system could be sustained and, furthermore, grow and thrive in
such circumstances?

\subsection{Fluidity versus Hierarchy?}\label{fluidity-versus-hierarchy}

Due to increasing social complexity, the future of global society
no longer resembles the past, and therefore our mental and formal
models lose their predictive power even in the short term
\citep{veitas_world_2015} resulting in an impression of chaos,
`crisis' and `a state of emergency'. While accelerating change and
information overload are the actual characteristics of the current
situation, the `state of emergency' is rather a subjective reaction
rooted in our prevailing worldview.\footnote{The concept of a worldview is instrumental for the conceptual 
framework of a social system which we are building in this paper and will be addressed in detail later (Section \ref{the-sense-making-and-a-worldview}).}

The prevailing worldview held today derives from the
\textit{Newtonian} worldview -- based on the concepts of reductionism,
determinism and objective knowledge \citep{heylighen_complexity_2006}.
Following this worldview we make sense of social reality by looking
for the existence of stable states in a social system. These states are
usually manifested as hierarchical relations among the system's elements,
participants or subsystems. Change is then conceptually understood as a
series of transitions between stable states.

\begin{figure}[H]
\centering
$\vcenter{\hbox{\includegraphics[height=1.3in]{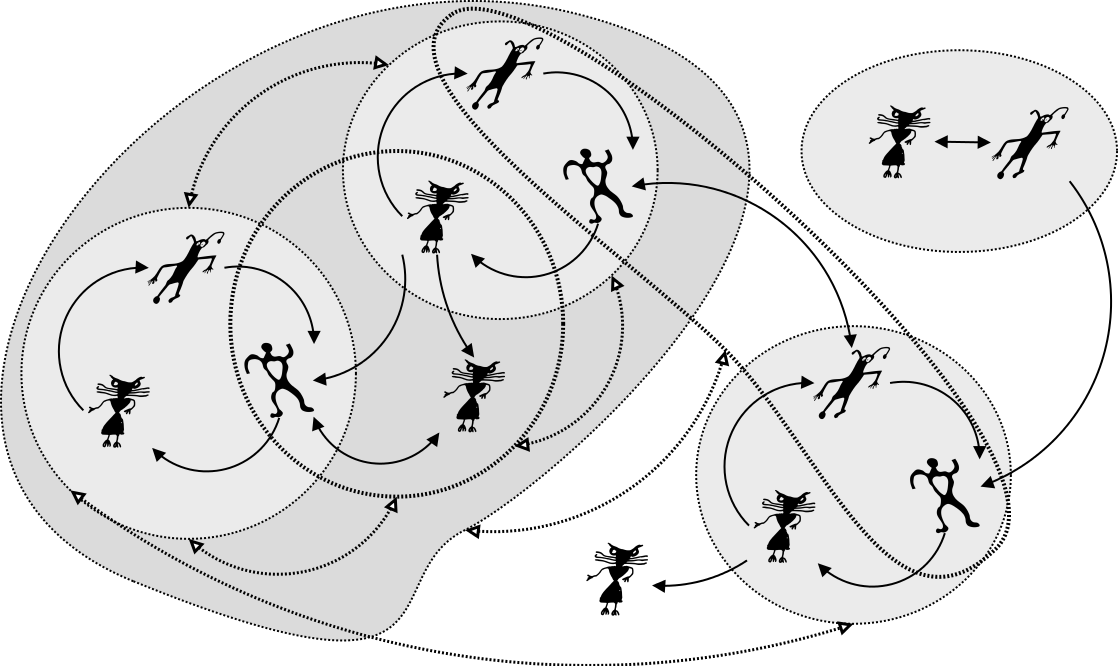}}}$
$\vcenter{\hbox{\includegraphics[height=1.3in]{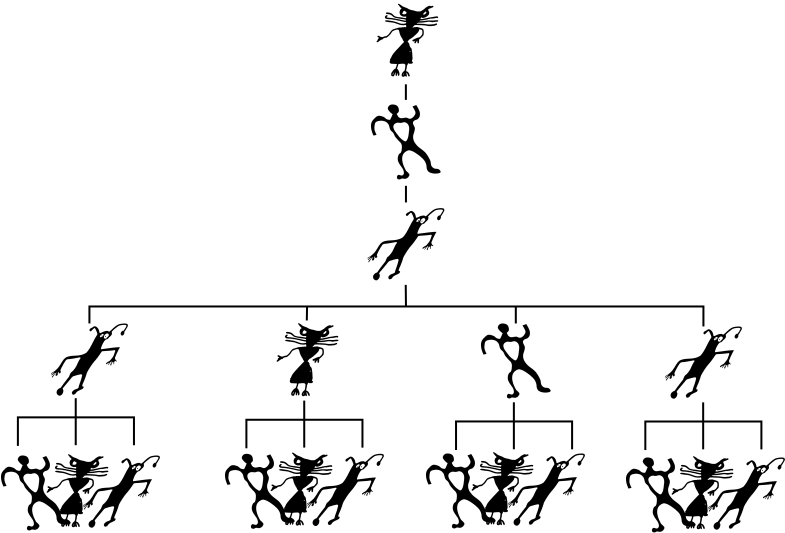}}}$
    \caption{\textbf{Fluidity or Hierarchy?}}
   \label{fig:chaos_vs_hierarchy}
\end{figure}

In other words, we are trying to mentally `stabilize' the increasingly
fluid and changing social system by finding more or less stable
hierarchical structures within it and then reflexively enforcing them
onto the system in the form of governance systems and institutions we
create. In terms of Figure \ref{fig:chaos_vs_hierarchy}, we are used
to thinking of social systems and global society as if having the
hierarchical structure on the right, while it actually resembles more
the image on the left. This discrepancy creates an impression that there
are no good models (or even worldviews) for understanding what is going
on.

In a situation of hyper-connectivity and accelerating change, the
`stabilization' operation becomes non-effective -- leading not only to
the impression of `crisis', `state of emergency' and ever growing
uncertainty, but also increasing tensions within and fragility of the
system. Seeing the global society as a dichotomy of `disorder versus
hierarchy' is counter-productive for understanding and governing it.

\subsection{A `Viscous' society}\label{a-viscous-society}

We therefore emphasize the view of the global society as a complex
system consisting of interacting subsystems at multiple scales. Nations,
states, religions, languages, local as well as international
institutions and governments, enterprises, philosophical schools and
academic institutions, fishing and golf clubs, families, persons and
pets are only a few examples of subsystems of the global society. While
social systems are neither completely fluid nor completely hierarchical,
we tend to see hierarchies in society, because: (a) temporary
hierarchical organizations do emerge and exist in it; and (b) it is
related to our `wired' tendency to search and see stable `coherent'
patterns in messy data.\footnote{Pattern recognition forms the basis for categorization and concept learning. 
Hierarchical organization is a particularly important way of organizing and relating concepts 
\citep{murphy_big_2004}, which is a necessary aspect of sense-making.}
But what we seem to de-emphasize are the fluid dynamics of a social
system which, we argue, form a more fundamental characteristic of the global
society than any observable stable state, which is never permanent.
Almost without exception however, contemporary governance structures are
organized hierarchically, which leads to the false impression that
society can be described and, moreover, governed, based solely on a
hierarchical model.

Certain social subsystems and units, such as linguistic dialects,
communities or religious beliefs, are fuzzy, overlapping and interactive
among themselves in a largely non-hierarchical manner. Others, usually
human-made systems, such as companies, armies and factories are
predominantly organized hierarchically.

Moreover, each social subsystem is constituted of a number of smaller
scale systems and each smaller scale system can be a member of more than
one subsystem at the higher level (Section
\ref{relations-between-scales}). For example, the same person can be a
member of a fishing and golf club and speak several languages.\footnote{See \citet{the_five_graces_group_language_2009} for the perspective on language as a complex adaptive system with a fundamentally social function.}
Furthermore, boundaries among certain subsystems, such as cultures or
philosophical schools, are far from being well defined or easily
definable . The conceptually coherent image of society is a
\textit{`viscous' system} -- combining different degrees of stability and
fluidity. Due to accelerating change, the level of `viscosity' of
society moves towards higher fluidity up to a point where the aspect
of hierarchical stability becomes hardly visible.

Therefore, we propose to approach any social system including global
society primarily as fluid, while treating observable hierarchical
structures as temporary `islands of stability' in an otherwise ever-changing 
social fabric. The next two sections introduce and discuss a
rich array of concepts and theoretical approaches integrated into the
framework of the Living Cognitive Society -- a fluid ecology of 
global society.

\section{Conceptual background}\label{conceptual-background}

The concept of the Living Cognitive Society integrates a number of
propositions brought forward by complexity science, cognitive science,
evolutionary theory, philosophy of individuation and becoming, and theory
of assemblages. In this section we briefly introduce each concept and
emphasize its influence and inspiration for the conceptual framework of
the Living Cognitive Society.

\subsection{Self-Organization in Complex Adaptive
Systems}\label{self-organization-in-complex-adaptive-systems}

The Living Cognitive Society in its most abstract definition is an
instance of a complex adaptive system (CAS). CAS are characterized by
complex patterns of behaviour which emerge from interactions among a
large number of component systems (agents) at different levels of
organization
\citep{chan_complex_2001,geli-mann_complex_1994,ahmed_overview_2005}.
The consequences of a huge number of interactions are most often
unpredictable due to their non-linear character. Still, interactions are
able to spontaneously coordinate among each other. Therefore, complex
adaptive systems are said to \textit{self-organize} instead of being
organized or designed.

Self-organization is the
\textit{appearance of structure or pattern without an external agent imposing it}
\citep{heylighen_science_2001}. Importantly, self-organization is caused
by a certain amount of disorder and fluctuations in the system --
formulated as principles of `order from noise' by Heinz von Foerster
and `order from fluctuations' by Ilya Prigogine
\citep{heylighen_science_2001}. These principles point to an important
understanding that fluidity, disorder, fluctuations and uncertainty are
not only `undesirable side effects' which should be minimized in a
complex adaptive system, but actually are \textbf{necessary} for it to
evolve, adapt and thrive. Therefore, a social system that can thrive in
uncertain environments, needs to reconcile a chaotic element in it -- a
crucial insight which we accommodate into the concept of the Living
Cognitive Society.

\subsection{Living and cognitive
systems}\label{living-and-cognitive-systems}

We see the virtue of combining the concepts of living and cognitive
systems to describe global society due to their potency to account
for emerging higher level coordination mechanisms within the system.

\begin{figure}[H]
\centering
    \includegraphics[width=0.65\textwidth]{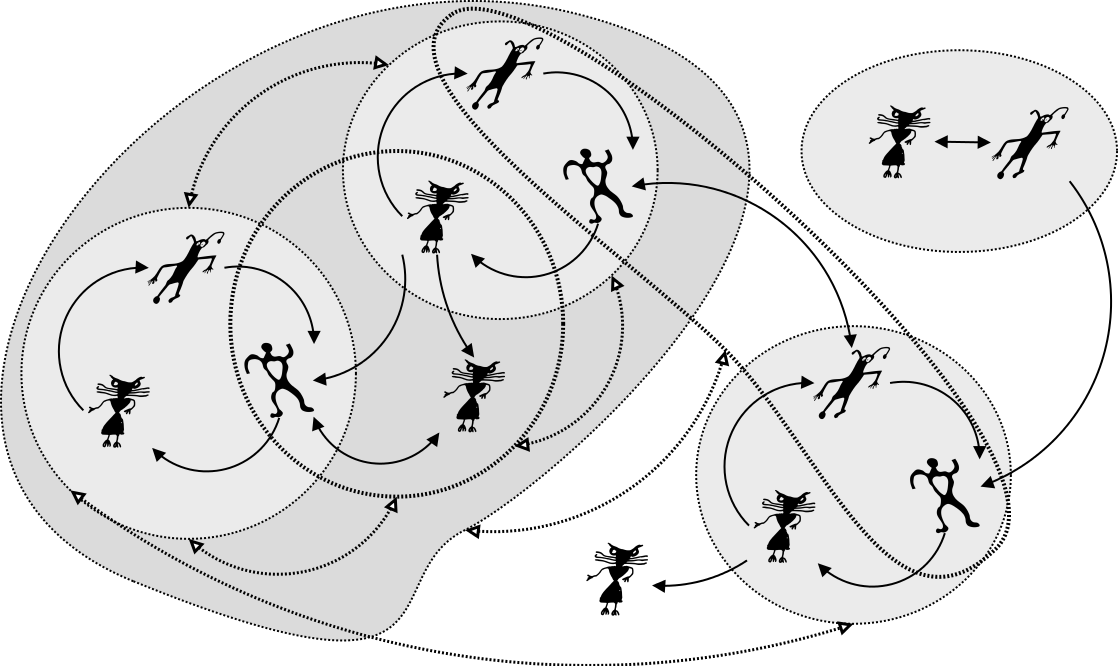}
   \caption{\textbf{The structure of global society.} }
   \label{fig:global_society}
\end{figure}

Therefore, we propose to analyse global society as a living system
\citep{miller_living_1975} which is also an ecology for other living
systems (Figure \ref{fig:global_society}). Examples of living systems
can be complex multi-cellular organisms exhibiting a high degree of
internal coordination (e.g.~vertebrates), but also loosely coordinated
organisms (e.g.~rhizomes and mycorrhizal networks). Clearly, the level
of coordination in the living system is a defining characteristic that
can bear disparate values.

Living and cognitive systems are categorized in the same equivalence
class \citep[p. 13]{maturana_autopoiesis_1980} or as closely related
\citep[p. 85]{luhmann_autopoiesis_1986}. Also,
\citet{miller_living_1975} treats society as a living system based on
the analysis of its properties.

In the context of global society, we are therefore faced with the
question of what the nature and dynamics of coordination are in a society
as a living system. Understanding of the close relationship between
living, cognitive and social is reflected in the name of the Living
Cognitive Society -- the central concept of this paper.

\subsection{Enaction}\label{enaction}

We largely subscribe to the research program of \textit{enaction}
\citep{stewart_enaction:_2010} for providing a conceptual framework of
self-organization in a living cognitive system. The enactive approach
treats cognition as an adaptive process of the interaction between an
agent and its environment. Importantly, it considers that the boundary
between an agent and its environment is constituted by these
interactions and largely defines an agent's \textit{identity}, whereas
identity of a self-organizing system is `generated whenever a
precarious network of dynamical processes becomes operationally closed'
\citep[p. 38]{stewart_horizons_2010}. Operationally closed networks of
processes are \textit{adaptively autopoietic} systems, i.e.~capable of
creating and sustaining themselves as well as continuously improving
their own conditions \citep[p. 50]{stewart_horizons_2010}
\citep[p. 78]{maturana_autopoiesis_1980}.

Simply put, living cognitive systems `have a say' in shaping the
tendencies that constrain and shape their own developmental dynamics and
effectively constitute their own identity. The enactive approach gives
us the understanding that these tendencies are not given from
\textit{outside} of the system, but are rather self-generated from the
interaction of the components \textit{within} the system.

\subsection{Sense-making and a
worldview}\label{the-sense-making-and-a-worldview}

The essence of the sense-making process is already encoded in the word
itself -- it is an active `making' of a `sense' or `meaning' by an
observer -- a cognitive agent. The concept does not overlook the fact
that sense-making is based on extracting information about observable
patterns in the system (the world, self and others) being perceived.
But, at the same time, it emphasizes that it is the observer who decides
what the \textit{significant} patterns are from which to extract the data
about a system or phenomenon. Sense-making is rooted in the enactive
approach to cognition (Section \ref{enaction}) which puts the concept in
a larger context, first of all entailing the individuation of the very
agent which performs sense-making.\footnote{We employ the simplification of a well defined observer - observed distinction (i.e. agent - environment) at this point mostly for didactic purposes. Actually, the distinction between observer and observed itself individuates during the process of synthetic cognitive development (Section \ref{synthetic-cognitive-development}). For in-depth analysis of the individuation of agent-environment boundary, please refer to \href{http://arxiv.org/pdf/1505.06366.pdf}{Section 2} of \citep{weinbaum_open_2015}.} For an in-depth
definition of the sense-making concept, please refer to
\href{http://arxiv.org/pdf/1411.0159v2.pdf
\#subsection.1.3}{Section 1.3} of \citep{weinbaum_synthetic_2014}.

The process of sense-making begets a \textit{worldview}. Importantly,
the relationship between sense-making and a worldview is a reflexive one
-- an observer's worldview determines significancies which then
influence the sense-making process of the same observer. The concept of
a worldview is a rich and multi-dimensional one (see
\cite{vidal_what_2008,vidal_beginning_2014} for an in-depth discussion
and references). It can be understood as
\textit{a gestalt perception -- unique and integrated cognitive structure -- held individually or collectively in relation to self, others, society, and the cosmos at large}
\citep{markley_changing_1981,veitas_world_2015}. With respect to the
social system we are living in, each worldview includes our aspirations,
the views on `natural tendencies' and `trends' of the system, and related
possibilities for the future as well as approaches to the appropriate
modes of social governance. Each of these aspects is based on a
combination of sense-making perspectives which may be overlapping,
incompatible or even mutually exclusive. For example, individuals or
collectives may prefer exploration, growth and development of persons,
society and life in general, or, alternatively, stability, safety and
preservation. Often such preferences cannot be accommodated within a
single value system and represent different `points of view' of the same
phenomena.

The Living Cognitive Society is the multiplicity of interacting
embodiments of worldviews, representing different value systems and
points of view. In a `viscous' society (Section
\ref{a-viscous-society}), where no single value system or worldview can
be considered dominant or `objectively' better/best, the resilience and
growth of the global system depends on the mode of interaction among
many worldviews rather than on the properties of any one of them.

\subsection{Synthetic cognitive
development}\label{synthetic-cognitive-development}

The theory of cognitive development posits identifiable patterns of
human cognitive development, which are described as developmental
\emph{stages} \citep{piaget_genetic_1971} or \emph{truces}
\citep{kegan_evolving_1982}, usually ordered in predictable sequences.
Cognitive development theories generally describe an `evolution of
meaning' \citep{kegan_evolving_1982}: the recursive subject and object
relationship when the subject of a previous stage, becomes an object
during the next stage. The process is not linear, but rather is
manifested through sequences of integration and disintegration of
cognitive structures (i.e.~developmental truces).

\begin{figure}[h]
\centering
\includegraphics[width=1\textwidth]{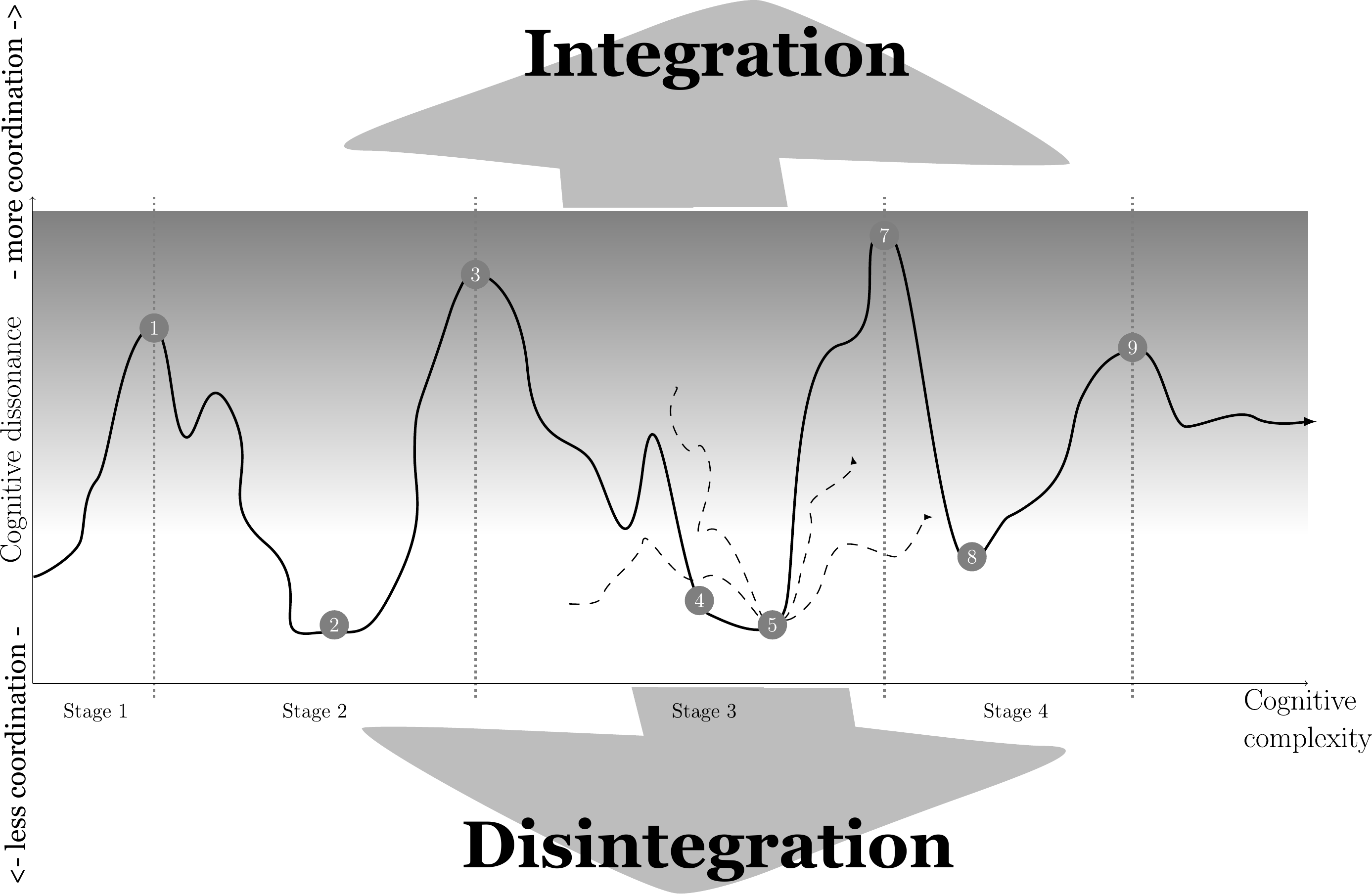}
   \caption{\textbf{A general scheme of synthetic cognitive development} qualitatively visualizing the process of increasing cognitive complexity as a variation of internal coordination levels within the cognitive system. The bold curve represents a possible development trajectory; Circles with numbers represent states of development, arbitrarily chosen for illustration.  States (1), (3), (7) and (9) mark high cognitive dissonance states where the system has the highest possibility of `choice' between alternative developmental trajectories.  Dashed lines are drawn at stage (7) to illustrate multiple possible trajectories that are actually present at every point along the developmental trajectory.  States (2),  (4),  (5) and (8) mark stable periods when the operation of a cognitive system is constrained.  Stages 1, 2, 3 and 4 on the horizontal axis illustrate cognitive development stages as described by the developmental psychology representing punctuated manner of increase in cognitive complexity. The process is reinforced by the interacting / alternating forces of integration and disintegration.}
   \label{fig:scheme_of_development}
\end{figure}

In \citet{weinbaum_synthetic_2014} we generalize the process of
cognitive development to all classes of living cognitive systems
(i.e.~humans, societies, artificial intelligences) and call it 
\textit{synthetic cognitive development}. We define synthetic
cognitive development in terms of the variability of the level of
internal coordination (Figure \ref{fig:scheme_of_development}) within a
complex adaptive system -- leading to the higher cognitive complexity of
a living cognitive system.

The generalization is achieved via a series of observations. First,
human cognitive development is largely driven by a
\textit{cognitive dissonance} -- a measure of subjective psychological
uncertainty.\footnote{\href{http://arxiv.org/pdf/1411.0159v2.pdf}{Section 1.4} in \citep[p. 6]{weinbaum_synthetic_2014}.}
Subjective uncertainty is nothing other than ignorance about the state
of the system and is related to the level of unpredictability of the
system's behaviour. \textit{Entropy} is used for measuring such
ignorance about the state of a system \citep{ben-naim_entropy_2012}.
Therefore, cognitive dissonance can be understood as the level of
entropy of a cognitive system -- a level of subjective uncertainty.\footnote{The entropic brain hypothesis by \citet{carhart-harris_entropic_2014}, showing that the entropy of brain activity is associated with mental
states, supports this observation.} Finally we observe that the level of
entropy/uncertainty in a cognitive system is consonant with the level of
disorder in a complex adaptive system. The concept of synthetic
cognitive development therefore operationalizes `order from noise' and
`order from fluctuations' principles describing self-organization
dynamics of complex adaptive systems (Section
\ref{self-organization-in-complex-adaptive-systems}). We apply these
general principles of cognitive systems' development for understanding 
global society. It allows us to start describing the interaction of
processes of integration (i.e.~leading to more order) and processes of
disintegration (i.e.~leading to more fluidity) and their primary role in
the self-organizing dynamics of the Living Cognitive Society.

\subsection{Theory of individuation}\label{theory-of-individuation}

The philosophy of individuation by Gilbert Simondon\footnote{Refer to  \href{http://arxiv.org/pdf/1411.0159v2.pdf}{Section 2.1} in \citep[p. 13]{weinbaum_synthetic_2014} for a brief philosophical introduction.}
is the ontological foundation of the conceptual background described in
this section. The theory of assemblages and the notion of tranduction --
central concepts required for understanding the workings of the Living
Cognitive Society -- are direct descendants of the theory of
individuation. Simondon opposes the hylomorphic schema which posits the
dichotomy of form and matter: he sees the form, the matter, the objects
and the relations among them individuating together without any primary
principle defined prior to this individuation. In the context of the
present paper we see this principle particularly useful for
conceptualizing the social reality, largely made of relations among
social actors. The theory of individuation allows us to reconcile social
structures and subsystems with social processes, which is a conceptual
core of the Living Cognitive Society:

\begin{quote}
The relation is not an accidental feature that emerges after the fact to give the substance a new determination. On the contrary: no substance can exist or acquire determinate properties without relations to other substances and to a specific milieu. To exist is to be connected. This philosophical proposition allows Simondon to establish the scope of his project: to reconcile being (\textit{l'\'etre}) and becoming (\textit{le devenir}) \citep[p. 77]{pascal_chabot_philosophy_2013}.
\end{quote}

Most importantly, Simondon's theory of individuation, while being an
abstract ontological framework, at the same time promotes what can be
called `concretization' -- the explanation of the emergence of
observable and graspable objects and relations in the social or
socio-technological reality, as well as the relationships between them. In
other words, `concretization' allows us to approach the emergence of
order from noise in an abstract way, as well as to apply the concept to the
specific system - the Living Cognitive Society.

\subsection{Theory of assemblages}\label{theory-of-assemblages}

The theory of assemblages was introduced by
\citet{deleuze_thousand_1987} and further modified and developed by
\citet{delanda_new_2006}. Our usage of the theory and its concepts is
motivated by a number of reasons:

\begin{itemize}

\item
The theory of assemblages provides an avenue for conceptualization of a generative model of individuation.
At its original level of abstraction, the theory could be seen as a direction towards formulating mechanisms of the process of \emph{becoming}, i.e. emergence of objects, systems and subsystems and their relations from initial noise and disorder.

\item
 Furthermore, \citet{delanda_new_2006} has developed the theory as a philosophical  framework explaining the emergence of scalable social entities such as personal networks, social organizations, markets, cities, nation  states, etc.
 General premises and concepts offered by it are  broadly applicable to the study of coalitions of cognitive agents and living systems, especially in cases of heterogeneous and hybrid populations of human and non-human agents \citep{weinbaum_framework_2012}.

\item
As will become clear later, we view any entity that can be assigned with certain degree of autonomy, intentionality and identity as a cognitive agent.
We maintain that such a view, partially inspired by \citet{latour_aramis_1996}, is reasoned and informative with respect to the global information society, especially considering its future developments in the short and medium-term \citep{veitas_world_2015}.
\end{itemize}

Furthermore, assemblage theory builds on the distinction between
relations of interiority and exteriority, which relates closely to the
distinction between subject and object in the framework of synthetic
cognitive development. It also develops the concepts of
territorialization and deterritorialization, which we reformulate as
processes of integration and disintegration within the framework of
synthetic cognitive development.

\begin{quote}
One and the same assemblage can have components working to stabilize its identity as well as components forcing it to change or even transforming it into a different assemblage. In fact one and the same component may participate in both processes by exercising different sets of capacities. \citep[p. 12]{delanda_new_2006}
\end{quote}

\subsubsection{Territorialization and
deterritorialization}\label{territorialization-and-deterritorialization}

The notion of interaction between processes of
\textit{deterritorialization} and \textit{territorialization} 
originated from \citet{deleuze_anti-oedipus:_1983,deleuze_thousand_1987},
first in the context of socio-economics of production, and then with
relation to the dynamical systems theory and self-organizing material
systems. \citet{delanda_new_2006} applies the concept when developing
the theory of assemblages as one of the dimensions / axes along which
the specific assemblage is defined. This dimension delineates variable
processes in which components of a system become involved and which
either stabilize the identity of an assemblage, by increasing its degree
of internal homogeneity and sharpness of its boundaries, or destabilize
it. The former is referred to as a process of
\textit{territorialization} and the latter as a process of
\textit{deterritorialization} \citep{delanda_new_2006}. The main
mechanism of territorialization is the formation of habitual repetition,
providing the assemblage with a stable identity, while that of
deterritorialization is the breaking of habits, effectively influencing and
changing its identity \citep[p. 34]{smith_gilles_2013}.

\subsubsection{Relations of interiority and
exteriority}\label{relations-of-interiority-and-exteriority}

The need to define the concept of relations of interiority and
exteriority comes from the notion of a scalable system -- a multiplicity
of recursively nested populations of heterogeneous assemblages which
themselves consist of populations of yet lower level elements. Relations
of interiority are defined as relations between lower-scale elements
\textit{within} the boundaries of an assemblage. Relations of
exteriority are relations among elements \textit{across} the boundaries
of an assemblage -- i.e.~with the elements of other assemblages in a
population.

From the perspective of synthetic cognitive development, the distinction
between relations of interiority and exteriority is fluid, because, due
to the processes of integration (territorialization) and disintegration
(deterritorialization) the boundaries of assemblages are ever-changing:

\begin{quote}
Already at the level of physical beings, {[}..{]} interiority and exteriority are not substantially different; there are not two domains, but a relative distinction; because, insofar as any individual is capable of growth, what was exterior to it can become interior \citep[p. 43]{combes_gilbert_2013}.
\end{quote}

The relation of the theory of assemblages to the framework of synthetic
cognitive development lies primarily in the conceptual understanding of
individuation and becoming as the interaction among processes of
integration and disintegration mediated by temporary structures emerging
within a cognitive system. The actual mechanism of this interaction is
unveiled by the concept of transduction, which is introduced next.

\subsection{Transduction}\label{transduction}

One of the most significant innovations in Simondon's theory of
individuation is the concept of \textit{transduction} -- the abstract
mechanism of individuation. Transduction lies at the basis of the
process of interaction between structure and dynamics of the Living
Cognitive Society. For the purposes of this paper we single out two
important aspects of the concept -- \textit{metastability} and
\textit{progressive determination}.\footnote{For in-depth introduction to the concept of transduction please refer to \href{http://arxiv.org/pdf/1411.0159v2.pdf}{Section 2.4} in \citep[p. 11]{weinbaum_synthetic_2014} and \href{http://arxiv.org/pdf/1505.06366v2.pdf}{Section 3.4} in \citep[p. 11]{weinbaum_open_2015}}

\subsubsection{Metastability}\label{metastability}

\begin{figure}[H]
\centering
\begin{minipage}[b]{0.7\textwidth}
\centering
\includegraphics[width=0.6\textwidth]{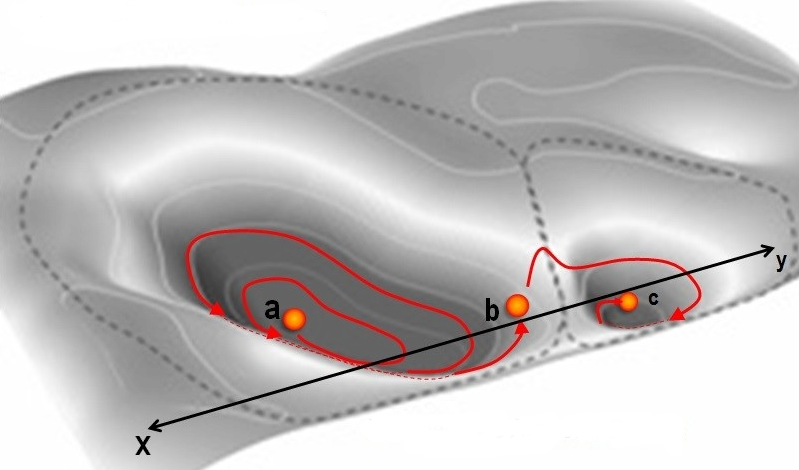}
\caption{\textbf{`Classical' metastability:} A metastable state of a weaker bond (a), a transitional `saddle' configuration (b) and a stable state of a stronger bond (c). Adapted from \citet{harrison2013building}}
\label{fig:metastability}
\end{minipage}
\end{figure}

The `classical' concept of metastability is mostly used to describe a
far-from-equilibrium complex system in terms of its movement in a
state-space. A metastable system can be seen as having a state-space
with many basins of attraction. Most of the time it is `being stuck' in
`shallow' attractors which may or may not be the state of system's least
energy. A metastable system can be easily perturbed, in which case it
moves over the border of one basin of attraction to another (Figure
\ref{fig:metastability}). A classical example of a physical system at a
metastable state is water at \(0\,^{\circ}{\rm C}\) temperature. If
water is still, it stays in a liquid state (even below the temperature
of \(0\,^{\circ}{\rm C}\)), but if it is perturbed by vibration, it
collapses into the state of ice.

What we add to the `classical' understanding of the metastability is the
fluidity of the state-space itself, meaning that it can adapt or otherwise become
perturbed. The concept of metastability within a fluid
state-space provides a concrete notion of what in the theory of
individuation is referred to as the \textit{pre-individual} -- a
seemingly disordered state from which an identifiable and observable
system may emerge.

The extended concept of metastability allows us to see the connection
between the theory of individuation and complex adaptive systems. The `order
out of chaos' principle can be intuitively grasped by imagining the
state space in Figure \ref{fig:metastability} being `shaken' by an
influx of additional perturbations. The energy from the noise increases
the probability of a system overcoming the `saddle' configuration and
ending up in another basin of attraction. In the case of a fluid state space,
the noise does not `shake' the state-space, but rather changes its
configuration. An important implication of this difference is that
in a fluid state-space we can relate the transformations of a state
space configuration to the movement of a system in it, without positing
an external source of noise or energy, as is usually done within the
framework of classical metastability. This operation is important for
understanding the mechanism of progressive determination.

\subsubsection{Progressive
determination}\label{progressive-determination}

Perhaps the most important aspect revealed in transduction is the
progressive co-determination of structure and operation. Progressive
determination can be seen as a chain of transformations where an
operation transforms a structure and a structure in turn transforms an
operation \citep[p. 11]{weinbaum_synthetic_2014}:

\[...S_1 \rightarrow O_1 \rightarrow S_2 \rightarrow O_2 \rightarrow S_3 \rightarrow ...\rightarrow O_{n} \rightarrow S_{n+1}...\]

\begin{itemize}
\item operation $O_i$ is a function which transforms one structure to another: $O_2 = S_i(O_1)$;
\item likewise, structure $S_i$ is a function which transforms one operation to another: $O_2 = S_i(O_1)$;
\item note that $S_1 \neq S_2$ and $O_1 \neq O_2$ -- they are  \textit{different} functions;
\item also note that $\rightarrow$ denotes the \textit{relations of dependency between} the transformations, so that every transformation depends on the full history of previous transformations.\footnote{I.e. it should \textit{not} be understood as a piping of inputs and outputs through the chain of immutable transformations.}
\end{itemize}

\section{Connecting the dots}\label{connecting-the-dots}

Let us now summarize our train of thought so far. First, in Section
\ref{the-current-situation-of-the-global-society} we asked the question of
whether our current social structures and, even more importantly, the
modes of thinking and making sense of the social reality which guides
the creation of these structures, are still valid in an era of ever-accelerating 
change. Our clear answer is `no'. We then ask a question of
\textit{what kind} of social system could thrive in these
circumstances. We argue that in order to conceive such a system we first
of all have to give up our prevailing modes of thinking about social
reality, specifically -- the assumption of supremacy of stability and
stable structures in it. The concept of the Living Cognitive Society --
the social system which we argue can remain resilient and thrive in 
circumstances of hyper-connectivity, accelerating change and
reflexivity -- combines influences from theories and conceptual
approaches discussed in Section \ref{conceptual-background}.

Drawing from complexity science and the concept of Complex Adaptive
Systems (Section \ref{self-organization-in-complex-adaptive-systems}) we
argue that in order to be resilient, the Living Cognitive Society has to
accommodate an element of disorder -- a necessary component of ecology
of interactions among heterogeneous social subsystems. This does not
mean that there should be no coordination, but rather emphasizes the
emergent nature of it. A clear example of such self-organized
coordination is to be found in living and cognitive systems (Section
\ref{living-and-cognitive-systems}). Processes driving emergence of
higher scale systems from the coordinated interaction of heterogeneous
elements of a population at a lower scale is the subject of the theory
of assemblages (Section \ref{theory-of-assemblages}). Most importantly,
the theory offers a concept of competing integrative and dis-integrative
processes (Section \ref{territorialization-and-deterritorialization})
leading to the emergence of higher order dynamics in a social system. We
further observe that the non-linear development of cognitive systems
happens via stages of integration and disintegration. Therefore,
cognitive development can be understood as a special case of formation
of assemblages driven by these processes. The concept of Synthetic
Cognitive Development generalizes insights from domains of cognitive
science and human cognitive development and applies them to the
development of social systems (Section
\ref{synthetic-cognitive-development}). Assemblages which are being
formed in the process of bottom-up self-organization are non other than 
`structures' which we observe in a social system. These observable,
yet often fuzzy, structures influence further dynamics of
self-organization in a system. The philosophical concept of transduction
(Section \ref{transduction}) provides an avenue for exploring and
understanding the mutual dependency between structure and dynamics in
the Living Cognitive Society.

Philosophy of individuation by Gilbert Simondon (Section
\ref{theory-of-individuation}) serves us as the conceptual glue for
integration of aforementioned disciplines and concepts via a carefully
constructed ontology where objects, their relationships, structures and
processes do not enjoy ontological primacy over one another but
\textit{individuate} via mutual interaction. In order for the
individuation of objects, their relationships and structures to take
place, the formation of boundaries between agents and environment has to
be explained, which is the emphasis of the enactive approach (Section
\ref{enaction}). The enactive approach treats cognition as the adaptive
process of agent-environment interaction. Both the theory of
individuation and the enactive approach deal with the abstract question of
how observable phenomena get determined from an indeterminate `fabric of
reality'.

These introductions to rich interdisciplinary sources which inspire our
thinking barely scratch their surfaces due to limitations of space and
scope of the paper. Yet our goal is not to fully describe these
theories, but rather provide the substantiation of characteristics of
the Living Cognitive Society concept, which we elaborate in the next
section.

\section{The Living Cognitive
Society}\label{the-living-cognitive-society}

The Living Cognitive Society is an ecology of emerging, interacting,
integrating and disintegrating cognitive systems at multiple scales
(Section \ref{structure-a-world-of-views}). This vision addresses
challenges of the current situation of the global social system (Section
\ref{the-current-situation-of-the-global-society}) and incorporates a
novel line of conceptual thinking (Section \ref{conceptual-background}).
Namely, it is a conceptual framework for conceiving the integration of
social institutions into the flexible, fluid and adaptable global
society operating in circumstances of uncertainty and change.

The Living Cognitive Society is described: (1) in terms of its scalable
structure -- \textit{A World of Views} (Section
\ref{structure-a-world-of-views}); (2) in terms of its dynamics -- the
process of \textit{Synthetic Cognitive Development} (Section
\ref{dynamics-synthetic-cognitive-development}); and (3) in the coupling and
interaction of structure and dynamics (Section
\ref{interaction-between-structure-and-dynamics}).

\subsection{Structure: A World of
Views}\label{structure-a-world-of-views}

Society is the vast ecology of interactions and communications among
agents -- more or less fuzzy integrated social assemblages and
institutions: nations, states, religions, cultures, companies and
governments, factories, academic institutions and families. If we
abstract from the concrete examples of social subsystems (i.e.~persons,
families, etc.) we can start regarding the \textit{scalable structure}
of the social system where interacting generic cognitive agents
(i.e.~individuals) assemble into cognitive agents at higher scales
(i.e.~organizations, cities), interactions among which create ecosystems
(i.e.~markets, communities, nations), which shape global society. In
a scalable system, every subsystem can be approached as an element of a
heterogeneous population which forms assemblages at a higher scale or,
alternatively, itself as an assemblage of a population of elements at a
lower scale. The adjacent scales of the system interact among
each other.

Abstracting further we observe that each social subsystem can be
understood as an embodiment of a certain worldview embedded in its own
unique social context. A \textit{worldview} is the integral system of
sense-making, incorporating cognitive and behavioural patterns which
govern social interactions of a system, embodying the worldview (Section
\ref{the-sense-making-and-a-worldview}). For example, individuals have
value systems, organizations and companies -- by-laws, cities --
regulations, states -- laws, etc., all of which are expressions of their
worldviews.

Social subsystems -- embodiments of worldwiews -- operate as
\textit{cognitive agents} with their own knowledge, competence, values,
goals and styles of behaviour. In a social context, a worldview can be
embodied as a person, a family, an organization or a company. But
actually any social subsystem with diverse levels of technological
involvement can be accommodated within this framework (Section
\ref{diversity}). By taking this perspective we enable ourselves to: (1)
approach the impact of technological developments to the social systems'
dynamics within the unified conceptual framework; (2) start describing
not only interactions at one scale (e.g.~persons with persons) but also
interactions between scales (e.g.~persons with nation states).\footnote{This aspect is central for discussion of a scalable system's dynamics and the concept of metastability (Section \ref{metastability}).}

A social system whose subsystems are abstracted from their specific
embodiments, i.e.~understood purely in terms of worldviews, is
\textit{A World of Views} -- a conceptual construct depicting society as
a multiplicity of interacting embodiments of unique, modular and
open-ended co-evolving worldviews. The construct of A World of Views was
developed as a philosophical framework and first used for describing a
futuristic socio-technological system by \citet{veitas_world_2015}. It
emphasizes an \textit{ecological} view of global society as a
superorganism \citep{heylighen_global_2002}, albeit having no single
locus of control. Here we apply the construct for the contemporary
global society and its near future -- related definitions are therefore
adapted for this context. For broader conceptual formulations we refer
the reader to the original article.

We started this paper by challenging the prevailing approach of looking
at social systems, in particular, and dynamic systems, in general, as series
of transitions between their identifiable stable states. The alternative
approach, named a `viscous society' (Section \ref{a-viscous-society})
approaches stable structures only as `islands of stability' in an ever-changing 
social fabric. We can look at the viscous society via the
metaphor of photography: a picture is `stable' only because it
captures the otherwise moving objects with the help of the short
exposure time. Yet it is possible to set a very long exposure and by
that make a picture where all fast moving objects (cars, people, sun,
stars, etc.) are unseen. In principle it is possible to set a long enough
exposure for the picture to be blank -- i.e.~not to capture \textit{any}
stable objects in it. The metaphor of photography illustrates how greatly
context-dependent is the property of stability of any given phenomenon we
observe. It also provides an intuition as to why in certain situations
(e.g.~when a photographer wants to capture the trajectories of planets in
the sky instead of the planets themselves) the pre-position of stable objects
does not allow to whole picture to be seen. We therefore now turn to
analysing the dynamics of social systems without this presumption.

\subsection{Dynamics: synthetic cognitive
development}\label{dynamics-synthetic-cognitive-development}

\subsubsection{Processes of integration and
disintegration}\label{processes-of-integration-and-disintegration}

Dynamics within the ecology of A World of Views, which structurally
describes the Living Cognitive Society, is based on interacting
processes of \textit{integration} and \textit{disintegration}. These
processes are the application of the theory of assemblages and generic
processes of territorialization and deterritorialization (Section
\ref{territorialization-and-deterritorialization}) to social systems.
Here we define these processes in the context of a scalable system,
i.e.~a system consisting of subsystems which themselves consist of
populations of yet another lower scale of `sub-sub systems' in a
recursive manner.

\begin{description}
\item[Integration] is a process which can happen locally or globally in a system and leads to higher levels of coordination among some elements of its population at any scale.
Clusters of elements which coordinate more strongly among themselves than with the rest of the population start forming an assemblage which, after reaching a certain level of internal coordination and resilience, can be identified as a newly formed subsystem with unique characteristics.\footnote{For the formal measures of coordination see also \citep[Appendixes A,B]{weinbaum_open_2015}.}
\item[Disintegration] is obviously the process of the opposite direction from integration: it leads to a lower level of coordination among elements of a given subsystem, ultimately reaching a level when a boundary between elements within the subsystem and elements outside dissipates -- i.e. the subsystem disintegrates.
\end{description}

Despite being always present, processes of integration and
disintegration are never symmetric: at every given moment either one is
stronger, giving rise to the complex dynamics of the living system in an
ecology of other living systems. The interplay between the processes of
integration and disintegration of variable strength and the importance
of this interaction in the growth of the cognitive system is captured by
the concept of synthetic cognitive development (Section
\ref{synthetic-cognitive-development}). Therefore, the maintenance of
the interaction of the processes of integration and disintegration in
the Living Cognitive System is instrumental for sustaining its
resilience and enabling open-ended development.

The lesson of complex adaptive systems is that processes of
disintegration (towards fluidity) are as important for the
self-organization of the system as processes of integration (towards
order). The awareness of such a balance is clearly missing from 
current approaches to social governance. As we have seen, the elements
of the social system are its subsystems -- institutions, organizations,
companies, businesses, governments, states. Therefore, trying to enforce
stability of social institutions -- something that we argue is the
prevailing paradigm of social governance -- makes the global system less
`alive' and therefore less adaptable and resilient, especially due to 
accelerating change, which requires ever-increasing elasticity.

\subsubsection{Relations between scales}\label{relations-between-scales}

\begin{figure}[H]
\centering
\begin{minipage}[b]{0.7\textwidth}
\centering
 \includegraphics[width=\textwidth]{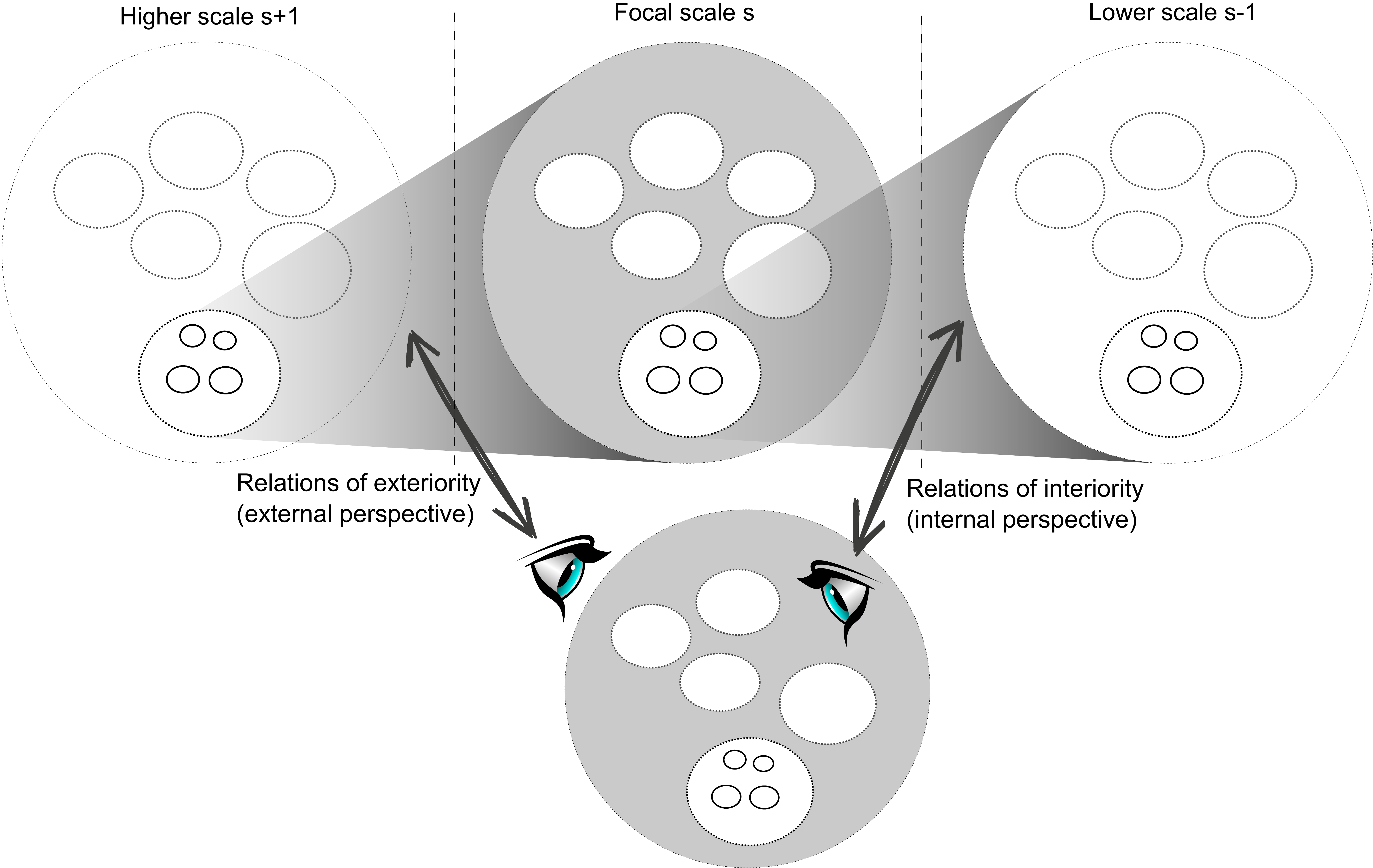}
\caption{\textbf{Relations between scales:} Lower scale $s-1$ is the population of heterogeneous elements from which a subsystem at focal scale $s$ integrates; higher scale $s+1$ is the heterogeneous population made of subsystems of scale $s$. From the perspective of the scale $s$ subsystem, the relation $S_{s+1} \rightarrow S_{s}$ is the relation of exteriority (i.e. external to the subsystem), the relation $S_{s} \leftarrow S_{s-1}$ is the relation of interiority (i.e. internal to the subsystem).}
\label{fig:relations_of_interiority_extriority}
\end{minipage}
\end{figure}

In a scalable system, every subsystem can be positioned at a
\textit{focal} scale \(s\) between higher \(s+1\) and lower \(s-1\)
scales (Figure \ref{fig:relations_of_interiority_extriority}). A lower
scale consists of a population of elements which integrate to a
subsystem at a focal scale; a higher scale consists of a population of
subsystems of the focal scale.

The processes of integration and disintegration at a focal scale are
driven by interactions at both higher and lower scales. Relations
between the focal scale and the higher scale are referred to as
\textit{relations of exteriority}; relations between the focal scale and
the lower scale -- as \textit{relations of interiority} (Section
\ref{relations-of-interiority-and-exteriority}). The above extends and
elaborates on a well known scheme of `upward' and `downward'
causation, which argues that `the whole is to some degree constrained by
the parts (upward causation), but at the same time the parts are to some
degree constrained by the whole (downward causation)'
\citep{heylighen_downcaus_1995,campbell1990levels}. Our extension
considers fluid boundaries between scales (i.e.~what is considered a
`whole' and `parts' in any specific situation) and the influence of the
processes of integration and disintegration on them. What it actually
means is that the distinction between interiority and exteriority, while
of utmost important for the operation of a cognitive agent as an
assemblage, itself gets individuated via the process of synthetic
cognitive development.

In the context of global society, the scheme depicted in Figure
\ref{fig:relations_of_interiority_extriority} offers a conceptual
outlook at how communities, nations and institutions emerge from
interactions and relationships of the population of heterogeneous
elements and further influence these interactions. It relates to the
notion of synthetic cognitive development by referring to the processes
which underlie integration and disintegration.

\subsection{Interaction between structure and
dynamics}\label{interaction-between-structure-and-dynamics}

In order to arrive at the complete picture of the ecology of the Living
Cognitive Society, we have to see how structure (Section
\ref{structure-a-world-of-views}) and dynamics (Section
\ref{dynamics-synthetic-cognitive-development}) inter-operate in a
metastable system. We propose to explain this interaction and, most
importantly, evolution of the structure-dynamics relations as a special
case of progressive determination (Section
\ref{progressive-determination}). We can recall that a metastable system is a
system which is permanently in a configuration other than the state of
least energy and having a fluid state space. Progressive determination
cannot be understood without the concept of metastability, as it
describes the very mechanism of fluid transformations of a metastable
state space (Section \ref{transduction}). Progressive determination of a
metastable system is the answer to the question which we asked at the
beginning of the article: `what kind of social system could be sustained
and, furthermore, thrive and develop in circumstances of
reflexivity, hyper-connectivity and accelerating change?'. That is, the
concept of transduction (Section \ref{transduction}) is applicable for
describing the mechanism of the operation of such a social system:

\begin{itemize}
\item First, it delineates how in a chain of transformations, every social structure is related to the momentum of immediate change happening in the system.
Likewise, every attempt to change the social system should be related to the current configuration of its social fabric.
\item Second, it points at the deeply rooted unpredictability of the process, which can lead to an either more or less integrated system.
\item The mechanism therefore implies a variety of possible configurations of a system.
\end{itemize}

The important message is that we cannot expect global society to be
resilient and growing by trying to artificially stabilize its structures
in the circumstances of all-pervasive accelerating change.

The structure of the Living Cognitive Society is a
metastable system applied to a social context and taking into account
the mechanisms of interaction among its internal subsystems. Every
observable structure in the Living Cognitive Society (nations,
institutions, families, communities, persons and their relationships)
should be understood as a specific state-space configuration of a
metastable system. As such, this observable structure (a) is temporary
and unstable, yet nevertheless (b) influences the further transformation and
evolution of global society. Furthermore, social institutions,
observable in the process of change, can have different and varying
levels of internal coordination -- i.e.~can be more or less integrated
depending on their level of cognitive dissonance (Section
\ref{fig:scheme_of_development}).

Armed with this conceptual model of the Living Cognitive Society we can
now describe how Information and Communication Technologies
(ICT) change global society in terms of their influence on the
process of progressive determination.

\section{The disruptive impact of information and communication
technologies}\label{the-disruptive-impact-of-information-and-communication-technologies}

\subsection{ICT and Distributed
Computing}\label{ict-and-distributed-computing}

Let us first define the two central concepts used in this section --
information communication technology and distributed computing.

\begin{description}

\item[Information and Communication Technology (\textit{ICT})]
acts as ``an umbrella term that includes any communication device or application, encompassing: radio, television, cellular phones, computer and network hardware and software, satellite systems and so on, as well as the various services and applications  associated with them''.
A complementary yet somewhat more modern term is \textit{cyberspace} -- a communication medium over computer networks, created by ICT.
Both terms refer to the same phenomenon, but with a clear difference between the emphasis on technical (ICT) and visionary (cyberspace) aspects.

\item[Distributed computing]
``arises when one has to solve a problem in terms of distributed entities such that each entity has only a partial knowledge of the many parameters involved in the problem that has to be solved'' \citep[p. v]{raynal_distributed_2013}.
Distributed computing entails not only the multiplicity of distributed interacting processes but also the fact that there is no single overarching process which can centrally integrate or control the interaction of this multiplicity.
Therefore, distributed computation embraces uncertainty and non-determinism and offers a computational perspective to complex adaptive systems.
\end{description}

We find the above computational perspective instrumental for describing
the processes within the Living Cognitive Society, especially with
relation to Information and Communication Technologies. According to this 
perspective the Living Cognitive Society is a distributed system 
consisting of a multiplicity of processes of progressive determination, 
modelled following the principles of distributed computing. Note, 
however, that while we borrow terms 
and concepts from computer science, we do not use them in a
strict formal sense. We rather use the concept
of distributed computation as a `lens for looking to the world'
\citep[p. xv]{moore_nature_2011} in order to show how seemingly simple
technical processes enabled by ICT can influence and disrupt the
dynamics of the ecology of global society in terms of the impact on
its structures (Section \ref{diversity}) and communication processes
(Section \ref{interactivity}).

\subsection{The mechanisms of
disruption}\label{the-mechanisms-of-disruption}

The disruptive impact of ICT on global society happens via the
cumulative effect of three mechanisms: (1) accelerating interaction
among the embodiments of worldviews in the Living Cognitive Society, (2)
multiplication and diversification of worldviews and their
embodiments, and (3) empowerment -- the increasing social power of
individuals. These mechanisms are reflexively interrelated: 
accelerating interaction stimulates the development of
social subsystems and catalyzes higher diversity, while diversity of
embodiments within the same medium brings about higher volumes of communication and
interaction. Empowerment of individuals is positively influenced by the
increasing fluidity of the global social system and at the same time
contributes to it. Each mechanism -- interaction, diversity or
empowerment -- taken separately, characterizes a long-standing tendency
of socio-technological development of human society which is not
directly related to information and communication technologies as they are currently understood. 
Yet, ICT contributes to the amplification of
all three mechanisms. Most importantly, it greatly facilitates a
positive feedback among them -- the
actual determinant of the disruption. Let us take a closer look at
each mechanism, considering their reflexive relations.

\subsubsection{Interactivity}\label{interactivity}

Dynamic interactions among social subsystems are triggered by 
factors of social complexity -- reflexivity, hyper-connectivity and
accelerating change (Section \ref{factors-of-social-complexity}) --
which cause an explosion of the amount of information flow within the
system. ICT enables, facilitates and supports this explosive
multiplication of information flow being exchanged among participants of
the system. Yet while the total amount of information in the system
grows, the ability of a single subsystem to process even a fraction of
this flow (note, that processing of information most importantly
includes selection for relevance) decreases. This phenomenon is called
\textit{information overload}.\footnote{It usually refers to human limits of information processing \citep{heylighen_complexity_2002}, but can be extended to any generic cognitive agent -- i.e. every system has certain limits of information processing. We therefore can apply the concept also for understanding the dynamics of hybrid populations of humans and technological artefacts.}
Due to it, any subsystem (individual, family, company and/or
country) is able to select for relevance and process an increasingly
minuscule fraction of the information available about the events
happening within the whole society. Therefore, the global system becomes
increasingly less graspable and predictable from the perspective of
\textit{any} subsystem. In order to successfully operate in such
environment, participants -- social subsystems -- have to increasingly rely
on the immediately available external information. 

ICT offers exactly that. Consider, for example, smart-phone usage for travelling 
in a city. Before internet based navigation technology became widespread (e.g. \href{http://maps.google.com}{GoogleMaps}; 
\href{https://www.apple.com/ios/maps/}{AppleMaps};  
\href{https://www.openstreetmap.org}{OpenStreetMaps}), people 
memorized their route trajectories, and used printed maps.
Note, that consulting a map application on a smart-phone is understood here as first of all an act of communication 
between two social subsystems -- a human user and the navigation technology. 
The technology thus made navigation much more efficient by allowing the user to rely 
on immediate external information obtained through interaction rather than internal 
representations of the route prepared in advance.
It allowed the user to consider quick changes of plans, transportation means and 
trajectories depending on the context (e.g. changed time of the meeting or 
delayed flight), which an is essential aspect of operation in an increasingly fluid environment.

Therefore, the nature of interaction and communication of subsystems
becomes relatively more important for the dynamics of the global social
system than individual properties, or the behavioural patterns of any single
participant, as well as the feasibility, predictive power or
accuracy of any model of the system. In the Living
Cognitive Society, these interactions among subsystems are perceived via the
concept of the relations of exteriority (Figure
\ref{fig:relations_of_interiority_extriority}), which, together with
relations of interiority, drive both the pace and nature of the cognitive
development of the system. In terms of the Synthetic Cognitive
Development (Figure \ref{fig:scheme_of_development}) this means that ICT
facilitates both integration and disintegration processes (Section
\ref{synthetic-cognitive-development}). Dynamic interaction between
integration and disintegration processes leads to accelerating
change in observable stable structures of global society via the
emergence of new subsystems and dissolution of the old ones up to a
point where it makes sense to say that all hierarchical structures
dissipate in favour of an ever-increasing fluidity of the system (Section
\ref{a-viscous-society}).

\subsubsection{Diversity}\label{diversity}

As we have seen, accelerating interaction via processes of
integration and disintegration brings about the emergence of more
diverse embodiments of worldviews in the ecology of the Living
Cognitive Society. Apart from facilitating the existing channels of communication, 
ICT also enables social subsystems -- the embodiments of the worldviews -- to
develop new ways of interaction within the same ecology. There is virtually no limit to the number of social identities and assemblages that can operate in a social system. We have already discussed how families, companies,
institutions and states can be approached as social subsystems of
various scales. The following examples illustrate different forms of
social subsystems starting with the pre-Internet era, where ICT had
little importance, and finishing with the ones for which cyberspace is
the basis of existence:

\begin{description}
\item[Dame Agatha Marie Clarissa Christie] was known by two `social identities':
\begin{inparaenum}[(1)]
\item Agatha Christie, who wrote 66 detective novels and 14 short story collections, and
\item Mary Westmacott who produced six romances;
\end{inparaenum}
\item[Nicolas Bourbaki] - sometimes called ``the greatest mathematician who never existed'' - was a group of 20\textsuperscript{th} century mathematicians which published a series of highly influential books under the collective pseudonym.
\item[Corporations] are treated as legal personalities -- non-human entities which are created by law with their own rights and responsibilities, not reducible to persons who are part of them.
\item[Wikipedia] is a famous example of how a trusted source of information can be created without trusted individuals involved in producing  it, and therefore can be approached as having distinct `social individuality'.
\item[Satoshi Nakamoto] is a person or group of people who created the Bitcoin protocol and reference software which started the `blockchain boom' with potentially wide disruptive results for the whole Internet ecosystem \citep{swan_blockchain:_2015}. The `actual' identity of Satoshi Nakamoto is unknown -- therefore it is a nice example of a social identity completely decoupled from the physical embodiment.
\item[Decentralized Autonomous Organization (DAO)] is a futuristic concept of an organization operating at an intersection of cyberspace and social reality.\footnote{The name of the concept has not yet stabilized, therefore it is also sometimes called \textit{fully automated business entity} or \textit{distributed autonomous corporation/company}.}
It is defined as a decentralized network of narrow-AI-type autonomous agents which performs an output-maximizing production function and which divides its labour into (a) computationally intractable tasks (which it can motivate humans to do) and (b) tasks which it performs itself \citep{babbitt_crypto-economic_2014}.
DAOs may play the role of business entities as well as non-governmental organizations without human management or even human involvement.
\end{description}

All the above are single and integrated social identities which in the
framework of the Living Cognitive Society are subsystems of the same
ontological status, i.e.~they are all embodiments of unique worldviews.

Cyberspace, as a digital medium, enables the easy creation of joint or
multiple identities not unlike the examples given above. We therefore
increasingly start to see social subsystems with the variable ratio of
human/technology involvement. Along these lines, the radical dissipation
of the difference between sociological and technological and the birth
of socio-technological can be best illustrated by the emerging concept
and technology of a decentralized autonomous organization. Therefore,
individuals or groups of people operating under pseudonyms, legal
corporations, croudsourcing projects, synthetic identities and DAOs can
be approached within the same framework of Living Cognitive Society as
social subsystems embodying diverse worldviews. ICT, being an enabler,
allows the embodiment and multiplication of worldviews, which would
be impossible without it.

Whether or not the dissipation of the difference between sociological
and technological reaches its radical levels, the influence of
information and communication technologies on the global social system
is profound in terms of increasing the number of subsystems within it
and fostering meaningful communications among them. These effects
will continue to increase the `birth' and `death' rates of
institutions, companies, states, communities, families, individual
social identities, etc., facilitating faster and larger data flows,
communication and interaction, inducing the higher fluidity of the
Living Cognitive Society.

\subsubsection{Empowerment}\label{empowerment}

Technology in general and information and communication technology in
particular, despite the immense possibilities it brings, is `only' an
enabler of different embodiments of worldviews and an amplifier of
their interactions. Therefore, while the development of a World of
Views and the way it will affect social life will be greatly enhanced
and enabled by ICT, the direction of the disruption will be determined
by the ``modes of the social inscription'' of these technologies
\citep{zizek_less_2013}.\footnote{For the philosophical / psychoanalytical speculations of the possible modes of social inscription, see \citep{zizek_what_2004}.}

A `real life' example of such possibilities could best be seen in the
`case' of the National Security Agency vs.~E. Snowden
\citep{poitras_citizenfour_2015} which illustrates a collision of two 
opposing social functions realized by the same technology: the first seeking total
surveillance and control; the other -- freedom and diversity of expression.
In the vocabulary of the Living Cognitive System, this is an example of
opposite worldviews embodied in the same technology. While
particularities of embodiment are important, the direction of
interaction is very much determined by the worldviews themselves. 

The case of `NSA vs. Snowden', as well as `U.S. vs.~WikiLeaks', illustrates
another important aspect of the future global society -- the greatly
increased capacities of individuals. A few years ago it would have been
unimaginable for one person or small group of individuals to `throw down
the gauntlet' to a powerful state agency. Empowerment of individuals
has profound systemic effects adding to the factors of social
complexity - by enabling persons or small groups to engage in
activities which can disrupt the whole global system.

\section{Governing the fluid society}\label{the-governance-of-fluid-society}

As we have seen, social subsystems are embodiments of worldviews which guide sense-making processes 
and, at the same time, are shaped by them (Section \ref{the-sense-making-and-a-worldview}).
The dynamics of the Living Cognitive Society is driven by these interacting worldviews via their embodiments.
This is the central corollary of our conceptual framework, which can by applied to the issue of governance of 
ecology of the Living Cognitive Society (Section \ref{distributed-social-governance}).

\subsection{Fluid society - A `digital' World of
Views}\label{fluid-society---a-digital-world-of-views}

The Living Cognitive Society will have a \textit{fluid identity} (or
rather it will be a \textit{fluid process}) emerged from an ecology of
interacting diverse embodiments of multiple worldviews. This fluid
identity will reflexively shape the underlying worldviews of its
constituting elements and subsystems.

We draw a clear parallel between Living Cognitive Society and the
perspective of the cognitive system as an ecology of interacting parts,
components, agents or thoughts
\citep{bateson_mind_2002,minsky_society_1988,dennett_consciousness_1993}.
The same perspective can be applied to society, individual humans, or any
social subsystem as a cognitive agent. No matter which social subsystem
we consider, it embodies certain perspectives of the environment that
surrounds it -- in other words, it `has' a worldview. As ICT enables
embodiment of images and worldviews of humans and their groups with
different degree of technology participation, we may see an explosion of
diversity of interacting identities -- digital, physical, `natural',
`artificial' and otherwise -- within the ecology of global society.
This ecology and its dynamics, rather than command-control hierarchies
which will increasingly become temporary and \textit{ad-hoc}, will
determine and accelerate the fluidity of the identity of the Living
Cognitive Society.

With the concept of the Living Cognitive Society we have provided our
answer to the central question of this paper: `what kind of social
system could be sustained and, furthermore, grow and thrive in 
circumstances of accelerating change?' Yet the framework of the Living
Cognitive Society raises a further question -- is it possible to coordinate or govern
such a system and what governance approaches will it require? The
elaborate answer to this question is beyond the scope of this work, yet
the governance of the Living Cognitive Society is essential to the
concept, and therefore the paper would be incomplete without touching major aspects
of it.

\subsection{Distributed Social
Governance}\label{distributed-social-governance}

The current hierarchical order of global society is often referred
to as `global governance' \citep{beauchamp_quiet_2015} or `post World
War II' structures of governance. At the core, this order amounts to the
complicated, yet highly hierarchically ordered network of governance
institutions at local, national and supranational levels. The ideal
system of global governance, following the prevailing perspective,
is a command and control hierarchy with a supranational institution
(e.g. the United Nations, or a `World Government') at its top. Yet there
is a clear perception that the `post World War II' structures are
failing. The response to the perceived risk of `fraying global
governance structures' usually amounts to `building better structures'
or `strengthening democratic institutions'
\citep{u.s._department_of_state_enduring_2015}, following the same
paradigm of global governance in terms of sustaining a command and control
hierarchy.

What we propose with the images of A World of Views and the Living
Cognitive Society is the shift of emphasis from the structures and
institutions to the very process of creation, adaptation and dissolution
of social subsystems at all scales of the global society. Furthermore,
the naturally distributed nature of the process -- meaning the absence of
central body or `trusted party' governing it -- should be embraced,
rather than fought with establishing global institutions or `world
governments' since, we maintain, no stable structure is able to
outweigh the factors of social complexity driving society towards
increasing fluidity. It is difficult to imagine such a system, which we
call a `distributed social governance', but the latest developments in
the ICT, especially Internet technologies, may provide important
insights and examples of the technological feasibility of this concept.

While developing the concept of the Living Cognitive Society, we have 
emphasized the ecological nature of the interactions among diverse social 
subsystems. This notion primarily comes to contrast the 
prevailing hierarchical view of social systems and their governance. Such a view is 
induced by the long-standing metaphor of the hierarchically controlled social organism. While scientific 
developments have already done away with the rigid mechanistic image of the organism 
\citep{heylighen_global_2002-1}, contemporary governance institutions still operate according to this obsolete 
paradigm. The Living Cognitive Society, in contrast, highlights the distributed and synergistic aspects of the 
social organism as an ecology, so balancing and complementing the rigid hierarchical view. This bias is 
well understood, taking into account the prevailing mode of how humans individually 
and society collectively relate to the world and themselves, but is not justified 
in terms of the pressing challenges of the current situation of 
the social system (Section \ref{the-current-situation-of-the-global-society}).

Therefore, the notion of distributed social governance most importantly relies on the 
explication of the close affinity of organism and ecology.\footnote{Recent research of microbiome points 
well to the immediacy of this relationship, suggesting that gut microbiota impact the 
cognitive function and fundamental behaviour patterns of humans, such as social 
interaction and stress management \citep{dinan_collective_2015,young_gut_2012}.} 
It propounds an approach to the Living Cognitive Society both in terms of the ecological perspective 
-- addressing the dynamics and fluidity of the system -- and organismic perspective -- illuminating the emergent 
coordination from synergistic interactions within it. No matter what kinds of technology will facilitate the 
future of governance, we believe in a distributed social governance that will be based \textit{not} on the design 
of optimal institutions, but rather on the fitting \textit{processes} which allow
for better coordination of the Living Cognitive Society.

\section*{Acknowledgements}

This work has been financially supported by the Global Brain Institute\footnote{\href{http://www.globalbraininstitute.org}{http://www.globalbraininstitute.org}}. It is based on the material presented at Summit 2015 of the International Society for Information Studies\footnote{\href{http://is4is.org/}{http://is4is.org/}}, conference track `The Global Brain and the Future Information Society'\footnote{\href{http://vienna2015.globalbraininstitute.org}{http://vienna2015.globalbraininstitute.org}}. The funding body had no influence on the content of the work, except support and encouragement for preparation and publication.

\printbibliography

\end{document}